\documentclass{paper}

\usepackage {plaatjes}
\usepackage {amssymb}
\usepackage {amsmath}
\usepackage {amsthm}
\usepackage [usenames] {xcolor}
\usepackage {float}
\usepackage {nicefrac}
\usepackage {wrapfig}
\usepackage {a4wide}
\usepackage [sort]{cite}

\newcommand {\mathset} [1] {\ensuremath {\mathbb {#1}}}
\newcommand {\R} {\mathset {R}}

\renewcommand {\L} {\mathfrak {L}}
\renewcommand {\S} {\mathfrak {S}}
\newcommand {\M} {\mathcal {M}}

\newcommand {\etal} {\textit {et al.}}
\newcommand {\eps} {\varepsilon}
\newcommand{\dif}{\mathrm{d}}
\newcommand{\cross}{\times}

\newtheorem {theorem} {Theorem}[section]

\newtheorem {lemma}[theorem] {Lemma}
\newtheorem {observation}[theorem] {Observation}
\newtheorem {corollary}[theorem] {Corollary}

\newenvironment{repeatlemma} [1]
{\noindent {\bf Lemma~\ref{#1}.}\ \slshape} {\normalfont}

\title {\Large Region-based approximation of probability distributions\\ (for visibility between imprecise points among obstacles)}

\author
{
  Kevin Buchin\thanks
  {
   Dept. of Mathematics and Computer Science,
   TU Eindhoven;
   \texttt{\{k.a.buchin, i.kostitsyna\}@tue.nl}.
  }
\and Irina Kostitsyna\footnotemark[1]
\and  Maarten L\"offler\thanks
  {
    Department of Computing and Information Sciences,
    Utrecht University;
    \texttt{m.loffler@uu.nl}.
  }
\and
  Rodrigo I. Silveira\thanks
  {
Dept. de  Matem\'atica \& CIDMA, Universidade de Aveiro, and
    Dept.\ Matem\`atica Aplicada II,
    Universitat Polit\`ecnica de Catalunya;
    \texttt{rodrigo.silveira@ua.pt}.
  }
}

\begin {document}

\maketitle

\begin {abstract}
  Let $p$ and $q$ be two imprecise points, given as probability density functions on $\R^2$, and let $\cal R$ be a set of $n$ line segments (obstacles) in $\R^2$.
  We study the problem of approximating the  probability that $p$ and $q$ can see each other; that is, that the segment connecting $p$ and $q$ does not cross any segment of $\cal R$.
  To solve this problem, we approximate each density function by a weighted set of polygons; a novel approach for dealing with probability density functions in computational geometry.
\end {abstract}

\tableofcontents

\section {Introduction}

Data imprecision is an important obstacle to the application of geometric algorithms to real-world problems.
In the computational geometry literature, various models to deal with data imprecision have been suggested.
Most generally,  in this paper  we describe the
location of each point by a probability distribution $\mu_i$ (for instance by a Gaussian distribution).  This model is often not worked with directly because of the computational difficulties arising from its generality.

These difficulties can often be addressed by approximating
the distributions by point sets.
For instance, for tracking uncertain objects a particle filter uses a discrete set of locations to model uncertainty~\cite{MDFW00}.
L\"offler and Phillips~\cite{lp-sfpspd-09} and J{\o}rgenson~\etal~\cite{jlp-gcip-11} discuss several geometric problems on points with probability distributions, and show how to solve them using discrete point sets (or \emph {indecisive} points) that have guaranteed error bounds.
More specifically, a 2-dimensional point set $P$ is an \emph {$\eps$-quantization} of an $xy$-monotone function $F$ (such as a cumulative probability density function), if for every point $q$ in the plane the fraction of $P$ dominated by $q$ differs from $F(q)$ by at most $\eps$.

Imprecise points appear naturally in many applications. They play an important role in
databases~\cite{DS04,ABSHNSW06,CM08,CG09,TCXNKP05,ACTY09,CLY09},
machine learning~\cite{BZ04}, and sensor networks~\cite{ZC04}, where a limited number of probes from a certain data set is gathered, each potentially representing the true location of a data point.  Alternatively, imprecise points may be obtained from inaccurate measurements or may be the result of earlier inexact computations.

Even though a point set may be a provably good approximation of a probability distribution, this is not good enough in all applications.
Consider, for example, a situation where we wish to model visibility between imprecise points among obstacles. When both points are given by a probability distribution, naturally there is a probability that the two points see each other.
However, when we discretize the distributions, the choice of points may greatly influence the resulting probability, as illustrated in Figure~\ref {fig:point-based}.

\eenplaatje {point-based} {Two pairs of point sets on opposite sides of a collection of obstacles. The green points can all see each other, whereas none of the blue points can.}

Instead, we may approximate distributions by regions.
The concept of describing an imprecise point by a region or shape was first introduced by Guibas \etal~\cite {gss-egbra-89}, motivated by finite coordinate precision, and later studied extensively in a variety of settings~\cite {gss-cscah-93,bs-ads-04,nt-teb-00,obj-ue-05,l-dicg-09}.

As part of our results we introduce a novel technique to represent the placement space of pairs of points that can see each other amidst a set of obstacles. 
We believe this technique is interesting in its own right.
For example, it can be applied to compute the probability that two points inside a polygon see each other, improving a recent result by Rote~\cite{r-dc-13} from $O(n^9)$ time to $O(n^2)$.

In this work we show how to use region-based approximation of point distributions to solve algorithmic problems on (general) imprecise points. In Section~\ref {sec:rba} we discuss several ways to do this. In Section~\ref {sec:vis}, we focus on a geometric problem for which previous point-based methods do not work well: visibility computations between imprecise points.

\section {Region-based approximation} \label {sec:rba}

Let $\M$ be a set of weighted regions in the plane, and let $w(M)$ denote the weight of a region $M \in \M$. Let $\M(p) = \{M \in \M \mid p \in M\}$ be the subset of $\M$ containing a point $p \in \R^2$. A set $\M$ defines a function $m (p) = \sum_{M \in \M(p)} w(M)$ that sums the weights of all regions containing $p$.

We say that $\M$ \emph {$\eps$-approximates} $\mu$ if the symmetric difference of the volumes under $m$ and $\mu$ is at most $\eps$; that is, if $\int_{p\in\R^2}|\mu(p)-m(p)| \le \eps$.
Figure~\ref {fig:region-approx-3d} illustrates the concept.

\eenplaatje {region-approx-3d} {A probability density function $\mu$ (yellow) can be approximated by a set of weighted regions $\M$, representing a function $m$ (purple).}

\paragraph {Additive or Multiplicative?}

To obtain a good set $\M$ that approximates a given density function, we make some observations.

Let $D \subseteq \R^2$ be a domain. We say $\M$ is a \emph {local additive $\delta$-approximation} on $D$ of $\mu$ if $|\mu(p) - m(p)| \le \delta$ for all $p \in D$. We say $\M$ is a \emph {local multiplicative $\delta$-approximation} of $\mu$ on $D$ if $(1-\delta) \mu(p) \le m(p) \le (1+\delta) \mu(p)$,  for all $p \in D$.

It is easy to verify that local multiplicative approximations imply global approximations:

\begin {observation}
  If $\M$ is a local multiplicative $\delta$-approximation of $\mu$ on $\R^2$, then $\M$ $\delta$-approximates $\mu$.
\end {observation}

However, there is a small problem: no finite $\M$ can be a local multiplicative approximation of many natural distribution (like Gaussians, for instance).
An earlier version of this document~\cite{previous,bkls-rbapd-14} mistakenly claimed that local \emph {additive} approximations imply global approximations.
This is not true: bounding the absolute distance between $m$ and $\mu$ at every point in the plane implies no guarantee on the error of these probabilities.
Figure \ref{fig:region-approx-2d} illustrates the difference between the two approaches.

\eenplaatje {region-approx-2d} {Illustration of the difference between additive (red) and multiplicative (blue) approximations of the same (1-dimensional) Gaussian function (yellow).}


Instead, the approach we will follow is to choose the number of regions and corresponding weights depending on the resulting volumetric errors.
Any probability distribution $\mu$ can be approximated in this way, but the total complexity of $\M$, i.e., the sum of the complexities of each of its regions, depends on various factors: the shape of $\mu$, the shape of allowed regions in $\M$, and the error parameter $\eps$. To focus the discussion, in this work we limit our attention to Gaussian distributions, since
they are natural and have been shown to be appropriate for modeling the uncertainty in commonly-used types of location data, like GPS fixes~\cite{horne2007analyzing,d-gnssa-07}.

\subsection {Approximation with Disks}
A natural way to approximate a Gaussian distribution by using a set of regions is by using concentric disks.
Thus, given a Gaussian probability distribution $\mu$, and a maximum allowed error $\eps$, we would like to compute a set $\M$ of $k$ disks that $\eps$-approximate $\mu$.
We may assume $\mu$ is centered at the origin, leaving only a parameter $\sigma$ that governs the shape of $\mu$, that is,
%
%
%
\[
\mu(x,y)=\frac{1}{2\pi\sigma^2}e^{-\frac{x^2+y^2}{2\sigma^2}}\,,
\]
or in polar coordinates,
\[
\mu(r,\theta)=\frac{1}{2\pi\sigma^2}e^{-\frac{r^2}{2\sigma^2}}\,.
\]
The function $\mu$ does not depend on $\theta$, therefore, in the following, we will omit it and write $\mu(r)$ for brevity.

\eenplaatje[scale=0.6] {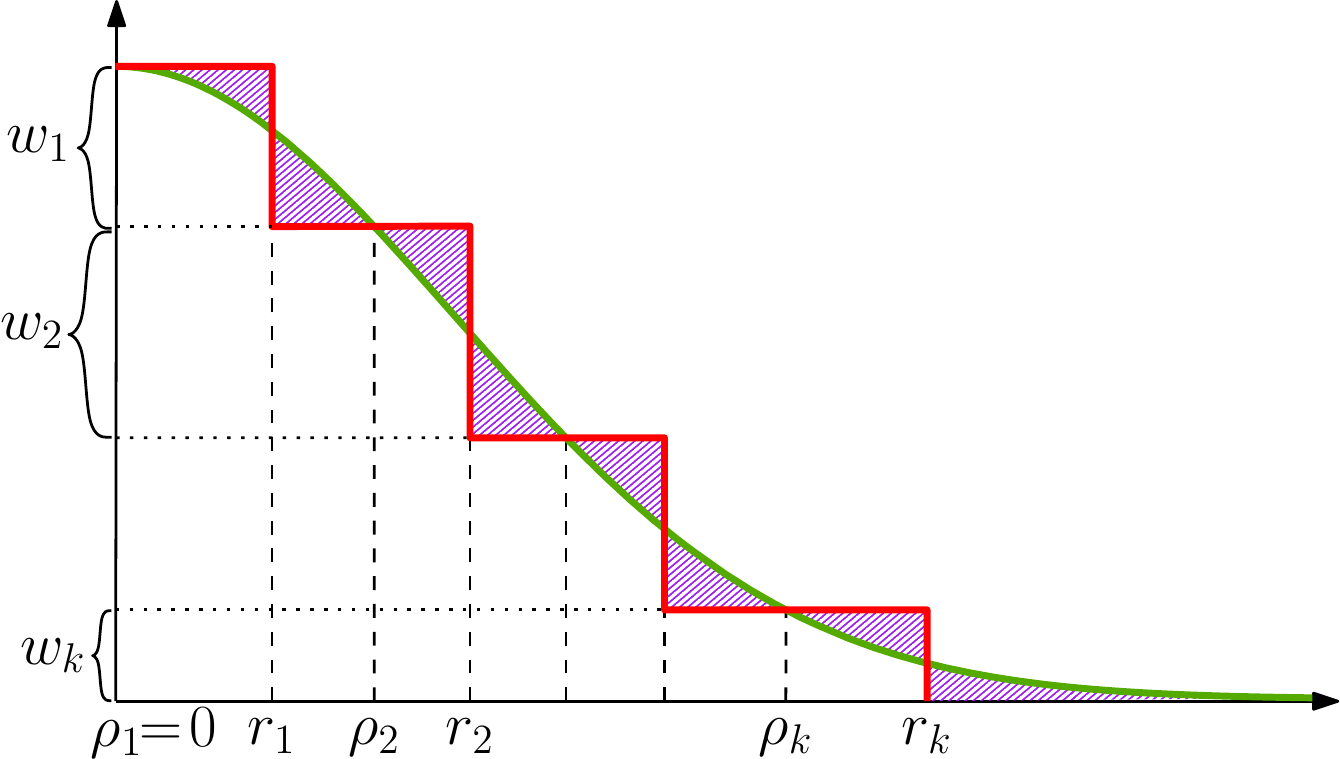} {Partial 2-dimensional cross-section illustrating the choice of radius ($r_i$) and weight ($w_i$) for the the $k$ regions in $\M$.
We use $\rho_i$ to indicate the $(i+1)$th radius where the approximation coincides with $\mu$.}

We are looking for a set of radii $r_1, \ldots, r_k$ and corresponding weights $w_1, \ldots, w_k$ such that the set of disks centered at the origin with radii $r_i$ and weights $w_i$ $\eps$-approximate $\mu$. We use these disks to define a cylindrical step function $\mu_{D}(r)$. 
Figure~\ref {fig:graph} shows a $2$-dimensional cross-section of the situation. 
Minimizing the volume between the step function and $\mu$, we obtain the following lemma:

\begin{lemma}\label{lem:disks}
Let $\mu$ be a Gaussian distribution with standard deviation $\sigma$. Let $k$ be a given integer. Then the minimum-error approximation of $\mu$ by a cylindrical step function $\mu_D$ consisting of $k$ disks is given by
\begin{equation}\label{eq:disks}
\begin{aligned}
r_{i}&=\displaystyle \sqrt{2\sigma^{2}\log{\frac{k(k+1)}{(k+1-i)^{2}}}}\,,\\
w_i&=\displaystyle \frac{1}{\pi\sigma^2}\frac{(k+1-i)}{k (k+1)}\,,
\end{aligned}
\end{equation}
where $i\in(1,\dots,k)$.
\end {lemma}

\begin{proof}
We sketch the proof idea here; the interested reader may refer to Appendix~\ref {app:disks} for the entertaining mathematical details.
To find the optimal weights, we introduce an additional set of parameters $\rho_1, \ldots, \rho_k$, where $\rho_i$ is the radius such that $\mu(\rho_i) = \sum_{j=1}^iw_j$, that is, it is those radii where the approximation and the true function intersect each other (see Figure~\ref{fig:graph}). We optimize over the $2k$ variables $r_i$ and $\rho_i$, by explicitly writing the symmetric difference as a sum of signed differences between $\mu$ and $\M$ over each annulus $(\rho_i, r_i)$ and $(r_i, \rho_{i+1})$. By equating the derivatives to $0$ we obtain the following useful identities:
\begin{align}
2\rho^2_{i}&=r^2_{i-1}+r^2_{i}\,,\label{eq:eq1}\\
2e^{-\frac{r^2_{i}}{2\sigma^2}}&=
\begin{cases}
\phantom{\Big|}e^{-\frac{\rho^2_{i}}{2\sigma^2}}+e^{-\frac{\rho^2_{i+1}}{2\sigma^2}} & \text{for }1\leq i<k\\
\phantom{\Big|}e^{-\frac{\rho^2_{i}}{2\sigma^2}} & \text{for }i=k\\
\end{cases}\label{eq:eq2}\,.
\end{align}
Further analysis yields the closed forms of expressions for $r_i$ (Equation~(\ref{eq:disks})) and $\rho_i$:
\begin{equation}
\rho_{i}=\sqrt{2\sigma^{2}\log{\frac{k(k+1)}{(k+1-i)(k+2-i)}}}\,.
\label{eq:rho}
\end{equation}

Substituting $w_i = \mu(\rho_i) - \mu(\rho_{i+1})$, we attain Equation~(\ref{eq:disks}), proving the lemma.
\end{proof}

Since the error allowed $\eps$ is given, we can use the expressions derived in the (full) proof of the previous lemma (in particular, Equation~(\ref{eq:volume-dif})) to find a value of $k$ such that the volume between the step function with $k$ disks and $\mu$ is at most $\eps$. 
This leads to the following result.

\begin {theorem}\label{thm:approx-disks}
Let $\mu$ be a Gaussian distribution with standard deviation $\sigma$.
Given $\eps>0$, we can $\eps$-approximate $\mu$ by a cylindrical step function $\mu_D$ that is defined by a set of
\[
k=\left\lceil\frac{1}{e^{\eps}-1}\right\rceil = O(1/\eps)
\]
weighted disks.
\end {theorem}
\begin{proof}
Using Equations~(\ref{eq:eq1}) and~(\ref{eq:eq2}) and Lemma~\ref{lem:disks}, Equation~(\ref{eq:volume-dif}) (see full proof of Lemma~\ref{lem:disks} in Appendix~\ref {app:disks}) can be simplified to
\[
F=\frac{r^{2}_{1}}{2\sigma^{2}}=\log{\frac{k+1}{k}}\,.
\]
Function $F$ gives the error of approximating the distribution function $\mu$ by the set of disks:
\[
\eps=F=\log{\frac{k+1}{k}}\,,
\]
and thus,
\[
k=\left\lceil\frac{1}{e^{\eps}-1}\right\rceil=O\left(\frac{1}{\eps}\right)\,.\qedhere
\]
\end{proof}

It follows that we can $\eps$-approximate a Gaussian distribution by using $O(1/\eps)$ disks.

\subsection {Approximation with Polygons}

The curved boundaries of the disks of $\mu_D$ make geometric computations more complicated. Therefore, next we consider approximating $\mu$ by a set of polygons. Computing a set of polygons of minimum total complexity is a challenging mathematical problem that we leave to future investigation. However, we can easily obtain a set of polygons at most twice as large as the minimum, by first computing a set of $k$ disks with guaranteed error $\eps$, then defining $2k$ annuli (two for each disk), and finally choosing $2k$ regular polygons that stay within these annuli. 
Figure~\ref {fig:isolines} illustrates this idea; since the relative widths of the annuli change, polygons of different complexity are used for different annuli. For each disk with radius $r_{i}$ we define two radii $r'_{i}$ and $r''_{i}$ by the following equations:
\begin{equation}
\begin{aligned}
\mu(r'_{i})&=\frac{1}{2}(\mu(\rho_{i})+\mu(r_{i}))\,,\\
\mu(r''_{i})&=\frac{1}{2}(\mu(r_{i})+\mu(\rho_{i+1}))\,.
\end{aligned}\label{eq:annuli}
\end{equation}

\tweeplaatjesbreed [height=2in] {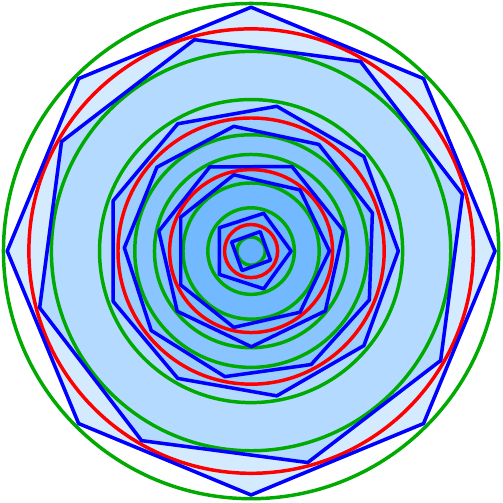} {graph-appx-complexity}
{ (a) A Gaussian distribution, given by isolines at $\eps$ levels (red), $2k$ annuli around each disk (green), and a set of polygons that can be used to obtain an approximation (blue).
  (b) We choose $2k$ regular polygons inscribed in annuli $\{r'_{i},r_{i}\}$ and $\{r_{i},r''_{i}\}$ with cumulative weights $W_{i}$ and $(W_{i}+W_{i+1})/2$, respectively.
}

%
Knowing the widths of the annuli we can calculate the total complexity of the approximation.

\begin {theorem}\label{thm:approximation}
 A Gaussian distribution with standard deviation $\sigma$ can be $\eps$-approximated by $O(1/\eps)$ polygons of complexity $O(1/\sqrt{\eps})$ each.
\end {theorem}
\begin{proof}
First, we compute a set of $k=\left\lceil\dfrac{1}{e^{\eps}-1}\right\rceil$ concentric disks by Equation~(\ref{eq:disks}) that approximate the distribution function $\mu$ with guaranteed error $\eps$. For each disk with radius $r_{i}$ we find two radii $r'_{i}$ and $r''_{i}$ from Equations~\ref{eq:annuli}. Then we choose $2k$ regular polygons that stay within annuli defined by pairs of radii $\{r'_{i},r_{i}\}$ and $\{r_{i},r''_{i}\}$ with weights $w_{i}/2$ each. These $2k$ polygons $\eps$-approximate the probability distribution function $\mu$. To prove this, we will show that this set of $2k$ weighted regular polygons approximates $\mu$ better than the cylindrical step function $\mu_D$ with $k$ disks. Consider all $r$ such that $\rho_i\leq r\leq \rho_{i+1}$. The value of $\mu_D$ is $W_i$ for $r\leq r_i$, and $W_{i+1}$ for $r>r_i$. The error of approximation of $\mu$ by $\mu_D$ at point $r$, therefore, is $W_i-\mu(r)$ for $r\leq r_i$, and $\mu(r)-W_{i+1}$ for $r>r_i$. Now consider the approximation of $\mu$ with the polygons. 
For all points within two annuli $\{\rho_i,r'_i\}$ and $\{r''_i,\rho_{i+1}\}$, the error of approximation of $\mu$ by the weighted polygons is exactly the same as by the disks (for these points, the weight of corresponding polygon is equal to the weight of the disks). 
For all points within two annuli $\{r'_i,r_i\}$ and $\{r_i,r''_i\}$, the error of approximation of $\mu$ by the weighted polygons is not greater than the error of approximation by the disks. 
For these points, the cumulative weight (that is, the value of the approximation) of the corresponding polygons equals the cumulative weight of the disks ($W_i$ for annulus $\{r'_i,r_i\}$, or $W_{i+1}$ for annulus $\{r_i,r''_i\}$), or is equal to $(W_i+W_{i+1})/2$. In the first case, again, the error of approximation of $\mu$ by the polygons in point $r$ is the same as the error of approximating it by disks. In the second case, using Equations~\ref{eq:annuli}, we conclude that the value of the approximation of $\mu$ by the polygons is closer to the true value of $\mu(r)$ than the one given by $\mu_D$ (refer to Figure~\ref{fig:graph-appx-complexity}). Therefore, the error of approximating $\mu$ by $2k$ weighted regular polygons is less than $\eps$.

\eenplaatje [scale=0.4] {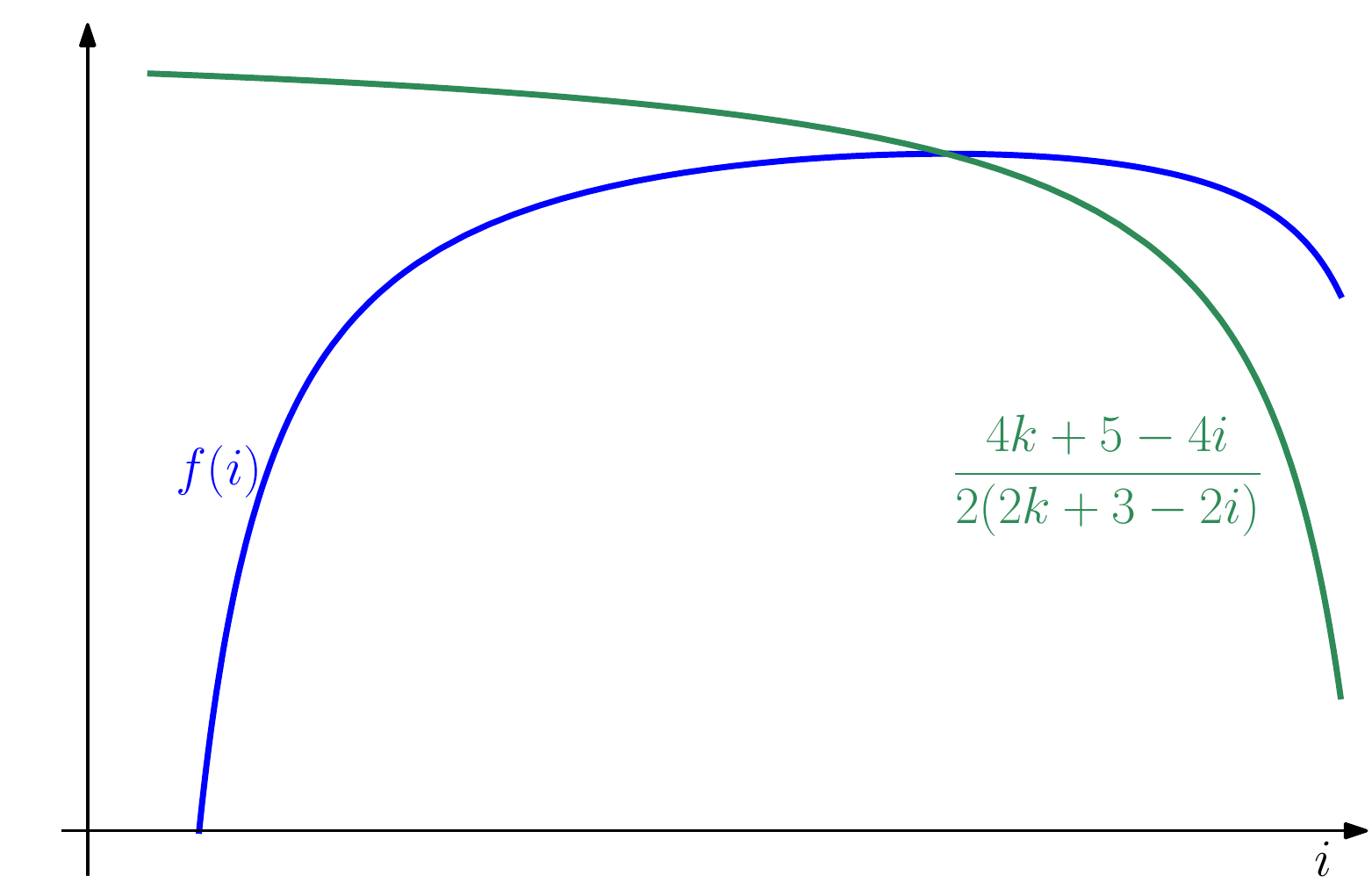} {Graphs of $f(i)$ and $\frac{4k+5-4i}{2(2k+3-2i)}$ intersect where $f(i)$ reaches its maximum.}
It remains to show that the complexity of each polygon is $O(\frac{1}{\sqrt{\eps}})$. The complexity of a regular polygon inscribed in an annulus depends only on the ratio of the radii. That is, given an annulus with inner radius $r'$ and outer radius $r$, we can fit a regular $\lceil\pi/\arccos\frac{r'}{r}\rceil$-gon in it. Similarly, given an annulus with inner radius $r$ and outer radius $r''$, we can fit a regular $\lceil\pi/\arccos\frac{r}{r''}\rceil$-gon. Consider the first case (the calculations for the second case are alike). First, derive from Equations~\ref{eq:annuli} the formula for $r'_i$:
\[
r'_i=\sqrt{2\sigma^2\log{\frac{2k(k+1)}{(k+1-i)(2k+3-2i)}}}\,,
\]
then the number of vertices $n'_i$ of the polygon inscribed in the annuli $\{r'_i,r_i\}$ is
\[
n'_i=\left\lceil\frac{\pi}{\arccos\frac{r'_i}{r_i}}\right\rceil=\left\lceil\frac{\pi}{\arccos{\sqrt{\frac{\log{\frac{2k(k+1)}{(k+1-i)(2k+3-2i)}}}{\log{\frac{k(k+1)}{(k+1-i)^{2}}}}}}}\right\rceil\,.
\]
Value $n'_i$ reaches its maximum when $\frac{r'_i}{r_i}$ is maximized. Consider $f(i)=\left(\frac{r'_i}{r_i}\right)^2=\frac{\log{\frac{2k(k+1)}{(2k+3-2i)(k+1-i)}}}{\log{\frac{k(k+1)}{(k+1-i)^2}}}$ as a continuous function of $i$, where $i$ is defined on interval $[1,k]$, differentiate it and solve the following equation:
\[
\frac{df}{di}=0\,.
\]
This leads to the following equation:
\[
2(2k+3-2i) \log{\frac{2k(k+1)}{(2k+3-2i)(k+1-i)}}-(4k+5-4i)\log{\frac{k(k+1)}{(k+1-i)^2}}=0\,.
\]
After dividing both sides of the equation by $2(2k+3-2i)\log{\frac{k(k+1)}{(k+1-i)^2}}$ (notice, that it is a non-zero value on interval $[1,k]$) we get
\[
\frac{\log{\frac{2k(k+1)}{(2k+3-2i)(k+1-i)}}}{\log{\frac{k(k+1)}{(k+1-i)^2}}}=\frac{4k+5-4i}{2(2k+3-2i)}\,.
\]
Notice, that the left-hand side of this equation is $f(i)$. Therefore, at maximum value of $f(i)$ it is equal to $\frac{4k+5-4i}{2(2k+3-2i)}$ (refer to Figure~\ref{fig:approximation-complexity}), and
\[
\max_{i}{f(i)}\leq\max_{i}\frac{4k+5-4i}{2(2k+3-2i)}=\frac{4k+1}{4k+2}\,.
\]
Thus, using the Taylor series expansion,
\[
n'_i\leq\left\lceil\frac{\pi}{\arccos{\sqrt{\frac{4k+1}{4k+2}}}}\right\rceil=2\pi\sqrt{k}+O\left(\frac{1}{\sqrt{k}}\right)=O\left(\frac{1}{\sqrt{\eps}}\right)\,.\qedhere
\]
\end{proof}

\section {Visibility between two regions} \label {sec:vis}

\tweeplaatjesbreed [scale=.6] {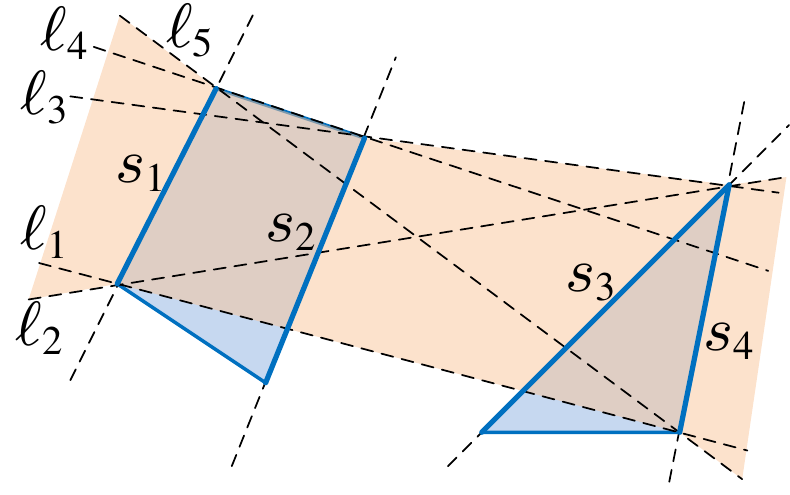} {dual-integral}
{ (a) Two polygons $P_{1}$ and $P_{2}$ in primal space. The orange region represents the set of  lines intersecting $P_{1}$ and $P_{2}$ through  $s_{1},s_{2},s_{3},s_{4}$.
  (b) Partition $L^{*}$ in  dual space. The orange cell corresponds to all lines in the primal space intersecting  $s_{1},s_{2},s_{3},$ and $s_{4}$.
}

Consider a set of obstacles $\cal R$ in the plane. We assume that the obstacles are disjoint simple convex polygons with $m$ vertices in total.
Given two imprecise points with probability distributions $\mu_{1}$ and $\mu_{2}$, we can approximate them with two sets of weighted regions ${\cal M}_{1}$ and ${\cal M}_{2}$, each consisting of convex polygons. 
For every pair of polygons $P_{1}\subset{\cal M}_{1}$ and $P_{2}\subset{\cal M}_{2}$, we compute the probability that a point $p_1$ chosen uniformly at random from $P_1$ can see a point $p_2$ chosen uniformly at random from $P_2$. We say that two points can ``see'' each other if and only if the straight line segment connecting them does not intersect any obstacle from $\cal R$. The probability of two points $p_{1}=(x_{1},y_{1})\in P_{1}$ and $p_{2}=(x_{2},y_{2})\in P_{2}$ seeing each other can be computed by the equation:
\begin{equation}
\label{eq:integral}
prob=\frac{\iiiint v(x_{1},y_{1},x_{2},y_{2}) \dif x_{1} \dif y_{1} \dif x_{2} \dif y_{2}}{\iiiint \dif x_{1} \dif y_{1}\dif x_{2} \dif y_{2}}\,,
\end{equation}
where $v(x_{1},y_{1},x_{2},y_{2})$ is $1$ if the points see each other, and $0$ otherwise.

To compute $prob$ we consider a dual space $\L$ where a point with coordinates $(\alpha,\beta)$ corresponds to a line $y=\alpha x-\beta$ in the primary space. We construct a region $\L^{*}$ in the dual space that corresponds to the set of lines that stab both $P_1$ and $P_2$. This region can be partitioned into cells, each corresponding to a set of lines that cross the same four segments of $P_{1}$ and $P_{2}$ (refer to Figure~\ref{fig:integral}).
The following follows from the fact that each vertex of $\L^{*}$ corresponds to a line in primary space through two vertices of $P_{1}$ and $P_{2}$.

\begin{observation}
Given two polygons $P_{1}$ and $P_{2}$ of total size $n$, the complexity of partition  $\L^{*}$ in the dual space that corresponds to a set of lines that stab $P_{1}$ and $P_{2}$ is $O(n^{2})$.
\end{observation}


\tweeplaatjesbreed [scale=.6] {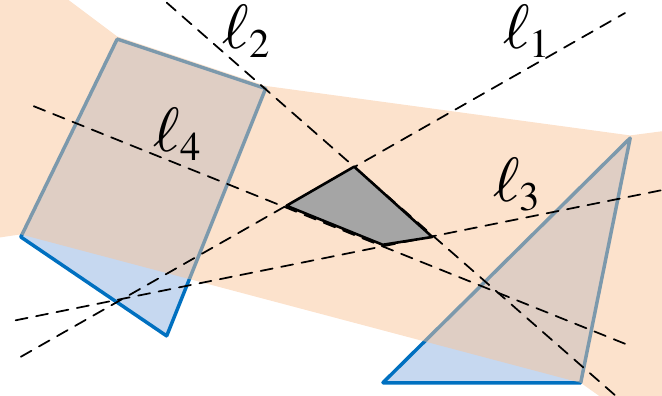} {dual-integral-obst}
{ (a) Primal space: polygons $P_{1}$ and $P_{2}$, and an obstacle between them.
  (b) Dual space: the ``hourglass'' shape $H^{*}$ (shown gray) in the dual space that corresponds to a set $H$ of all lines in the primal space that intersect the obstacle.
}

For each obstacle $h\subset {\cal R}$ we construct a region $H^{*}$ in the dual space, that corresponds to the set of lines that intersect $h$. $H^{*}$ has an ``hour-glass'' shape (refer to Figure~\ref{fig:integral-obst}). We now compute the subdivision $\L$ of the dual plane resulting from overlaying the partition $\L^{*}$ and the regions $H^{*}$. Since the objects involved are bounded by a total of $O(m+n)$ line segments in the primal space, $\L$ has complexity $O((m+n)^2)$.

First consider the case that $P_1$, $P_2$ and the obstacles are disjoint. We can assume that all obstacles lie in the convex hull of $P_1$ and $P_2$. Then a pair of points from $P_1$ and $P_2$ see each other exactly if the line through the points does not intersect an obstacle. Thus, we only need to identify the cells in $\L$ not intersecting any of the regions $H^{*}$, and integrate over these cells. Details on evaluating the integral for one cell are given in Section~\ref{sec:int}. Overall, this case can be handled in $O((m+n)^2)$ time.

Next, consider the case that $P_1$ and $P_2$ are disjoint but might intersect obstacles. Now we need to consider the length of each line segment from the last obstacle in $P_1$ to the boundary and from the boundary of $P_2$ to the first obstacle. We can annotate the cells of $\L$ with this information by a traversal of $\L$. Between neighboring cells this information can be updated in
constant time.
Thus, this case can be handled with the same asymptotic running time as the previous case.

As a third case, consider $P_1$ overlapping $P_2$ but with no obstacles in the overlap area. The computations needed remain the same as in the case of non-overlapping $P_1$ and $P_2$.
provided we actually evaluated this integral, we should now be able to compute the value in $O(n^2)$ time.

Finally, we consider the general case, in which obstacles might also lie in the overlap of $P_1$ and $P_2$.
In the cells of $\L$ that correspond to the overlap of $P_1$ and $P_2$ we now need to consider the sum of the lengths of each line segment between boundaries of obstacles.
If we simply traverse $\L$, maintaining the ordered list of intersected obstacle boundaries, then computing the sum of lengths in one cell requires $O(m)$ time, leading to a total running time of $O(m(m+n)^2)$.
Instead, we investigate the structure of the problem a little more closely.

\begin {lemma} \label {lem:nom}
  Let $P$ be a polygon, possibly with holes or multiple components, of total complexity $n$.
  Let $\S$ be the space of all maximal line segments, that is, segments which lie in the (closed) interior of $P$ but which are not contained in larger line segments that also lie in the interior of $P$.
  Then $\S$ has complexity $O(n^2)$.
\end {lemma}

\vierplaatjes {ms-primal} {ms-dual} {ms-stack} {ms-graph}
 { (a) A square polygon $P$ with two holes (obstacles).
   (b) The dual space $\L$ (cropped to a square). The colored area corresponds to all lines in $\L$ that intersect the square domain. The orange/yellow area corresponds to the lines that intersect the red obstacle; the blue/yellow area corresponds to the lines that intersect the green obstacle.
   (c) The space of maximal line segments $\S$. The purple layer are the segments that miss both obstacles (they extend from one end of the square to the other). The yellow layer are the segments that touch both obstacles. The red and orange layers are the segments that touch the red obstacle, but miss the green obstacle. The blue and green layers are the segments that touch the green obstacle, but miss the red obstacles. Vertical panels indicate which edges of layers are connected.
   (d) Schematic view of how layers are connected to each other.
 }

\begin {proof}
Line segments have four degrees of freedom, but the condition that they must be locally maximal removes two of them, so $\S$ is intrinsically two-dimensional. We may project $\S$ onto the set $\L$ of all lines (by extending each segment to a line), but this way we may map multiple segments onto the same line. However, we only map finitely many segments to a line. We can visualize this as a finite set of ``copies'' of (patches of) $\L$ above each other. Then, as we move (translate or rotate) our segment through $P$, it may split into two segments when we hit a vertex; this corresponds to one of the copies of $\L$ splitting into two copies. The ``seams'' along which the copies of patches of $\L$ are sewn together in $\S$ are one-dimensional curves, which correspond to the segment in $P$ rotating around (and touching) a vertex. The endpoints of these seams are points which correspond to segments in $P$ that connect two vertices.
Figure~\ref {fig:ms-primal+ms-dual+ms-stack+ms-graph} illustrates $P$, $\L$ and $\S$ for a small example.

Clearly, there can be at most $O(n^2)$ segments that connect two vertices in $P$, thus, there are only $O(n^2)$ vertices in $\S$. This does not immediately give the bound, though, since $\S$ is not planar. However, each vertex in $\S$ (corresponding to a pair of vertices in $P$) can be incident to at most two seams in $\S$: one that corresponds to a segment rotating around either vertex in $P$.
So, the total number of seams can also be at most $O(n^2)$.
Since a seam always connects exactly three patches, the total complexity of $\S$ is $O(n^2)$.
\end{proof}

If we apply Lemma~\ref {lem:nom} to our setting, then $P$ is the imprecise point with obstacles as holes, of total complexity $(n+m)$.
We arrive at the following intermediate result. In the next section, we show how to compute the probability for a given combinatorial configuration.

\begin {lemma}
  Given two polygons $P_1$ and $P_2$ of total size $n$ and obstacles of total complexity $m$, we can compute the probability that a pair of points drawn uniformly at random from $P_1 \cross P_2$ can see each other in $O((m+n)^2)$ time, assuming we can compute the necessary information within each cell.
\end {lemma}

\section {Computing the probability for a fixed combinatorial configuration} \label {sec:int}

For simplicity of presentation,
we assume that $P_{1}$ and $P_{2}$ are separable by a vertical line, and $P_{1}$ and $P_{2}$ are disjoint from $\cal R$. This will allow us to write the solution in a more concise way without loss of generality.

Consider line $\ell$, given by the equation $y=\alpha x-\beta$, that goes through two points $p_{1}(x_1,y_1)\in P_{1}$ and $p_{2}(x_2,y_2)\in P_{2}$. In the dual space, point $\ell^{*}$, corresponding to line $\ell$, has coordinates $(\alpha,\beta)$. Substitute variables $y_{1}$ and $y_{2}$ in Equation~(\ref{eq:integral}) with $\alpha$ and $\beta$:
$
(x_{1},y_{1},x_{2},y_{2})\leftarrow(x_{1},\alpha,x_{2},\beta)$,
where $\alpha(x_{1},y_{1},x_{2},y_{2})={y_2-y_1}/{x_2-x_1}$
and $\beta(x_{1},y_{1},x_{2},y_{2})=({x_{1}y_{2}-x_{2}y_{1}})/({x_2-x_1})$.
We can express the probability of two points, distributed uniformly at random in $P_{1}$ and $P_{2}$, seeing each other as
\begin{equation}
\label{eq:subst-integral}
prob=\frac{\iiiint v(x_{1},\alpha,x_{2},\beta)|J| \dif{x_{1}}\dif{x_{2}}\dif\alpha \dif\beta}{\iiiint |J|\dif{x_{1}}\dif{x_{2}}\dif\alpha \dif\beta}\,,
\end{equation}
where
\[
J=\det\left[\begin{array}{cc}
\frac{dy_{1}}{\dif\alpha} & \frac{\dif y_{1}}{\dif\beta} \\
\frac{dy_{2}}{\dif\alpha} & \frac{\dif y_{2}}{\dif\beta} \end{array}\right]=\frac{1}{\det \left[ \begin{smallmatrix}
\frac{\dif\alpha}{\dif y_{1}} & \frac{\dif\beta}{\dif y_{1}} \\
\frac{\dif\alpha}{\dif y_{2}} & \frac{\dif\beta}{\dif y_{2}} \end{smallmatrix} \right]}=x_2-x_1\,.
\]

The denominator of~(\ref{eq:subst-integral}) can be written as a sum of integrals over all cells of partition $L^{*}$ in the dual space:
\[
\sum_{C\subset L^{*}}\iint\limits_{C} \left( \int\limits_{X_{1}(\alpha,\beta)}^{{X_{2}(\alpha,\beta)}} \int\limits_{X_{3}(\alpha,\beta)}^{{X_{4}(\alpha,\beta)}}(x_{2}-x_{1}) \dif{x_{2}} \dif{x_{1}}\right) \dif\alpha \dif\beta\,,
\]
where $X_{1}(\alpha,\beta)$, $X_{2}(\alpha,\beta)$, $X_{3}(\alpha,\beta)$, and $X_{4}(\alpha,\beta)$ are the $x$-coordinates of intersections of line $y=\alpha x-\beta$ with the boundary segments of $P_{1}$ and $P_{2}$.

The numerator of~(\ref{eq:subst-integral}) can be written as a sum of integrals over all cells of partition $L^{*}\backslash\cup_{h} H^{*}$ in the dual:
\[
\sum_{C\subset L^{*}\backslash\cup_{h}H^{*}}\iint\limits_{C} \left( \int\limits_{X_{1}(\alpha,\beta)}^{{X_{2}(\alpha,\beta)}} \int\limits_{X_{3}(\alpha,\beta)}^{{X_{4}(\alpha,\beta)}}(x_{2}-x_{1}) \dif{x_{2}} \dif{x_{1}}\right) \dif\alpha \dif\beta.
\]

In Appendix~\ref {sec:appx-integral} we give a detailed case-by-case closed-form evaluation of the integrals. 
Since we integrate over constant-size subproblems, we obtain:

\begin {theorem} \label {thm:single}
Given two polygons $P_{1}$ and $P_{2}$ of total size $n$ and a set of obstacles of total size $m$, we can compute the probability that a point $p_1$ chosen uniformly at random in $P_1$ sees a point $p_2$ chosen uniformly at random in $P_2$ in $O((m+n)^2)$ time.
\end {theorem}

As an easy corollary, we improve on a result by Rote~\cite{r-dc-13}, who defines the ``degree of convexity'' of a polygon as the probability that two points inside the polygon, chosen uniformly at random, can see each other.

\begin {corollary}
 Let $P$ be a polygon (possibly with holes) of total complexity $n$. We can compute the probability that two points chosen uniformly at random in $P$ see each other in $O(n^2)$ time.
\end {corollary}

\section {Main result}
Combining Theorems~\ref {thm:approximation} and~\ref {thm:single}, our main result follows:

\begin {theorem}
  Given two imprecise points, modelled as Gaussian distributions $\mu_1$ and $\mu_2$ with standard deviations $\sigma_1$ and $\sigma_2$, and $n$ obstacles, we can $\eps$-approximate the probability that $p$ and $q$ see each other in $O(\sigma_1^{-2}\sigma_2^{-2}\eps^{-2}((\sigma_1^{-2}+\sigma_2^{-2})\eps^{-1}+n)^2)$ time.
\end {theorem}

\begin {proof}
  According to Theorem~\ref {thm:approximation}, we need to solve $O(\sigma_1^{-2}\sigma_2^{-2}\eps^{-2})$ individual problems.
  For each, we have $m = O((\sigma_1^{-2}+\sigma_2^{-2})\eps^{-1})$, so using Theorem~\ref {thm:single} we solve them in $O(((\sigma_1^{-2}+\sigma_2^{-2})\eps^{-1}+n)^2)$ time.
  This leads to
  $O(\sigma_1^{-2}\sigma_2^{-2}\eps^{-2}((\sigma_1^{-2}+\sigma_2^{-2})\eps^{-1}+n)^2)$  running time.
\end {proof}

\small
\textbf{Acknowledgments.}
K.B., I.K., and M.L.\ are supported by the Netherlands Organisation for Scientific Research (NWO) under grant no. 612.001.207, 612.001.106, and 639.021.123, respectively.
R.S. was funded by Portuguese funds through CIDMA and FCT, within project PEst-OE/MAT/UI4106/2014, and by FCT grant SFRH/BPD/88455/2012.
In addition, R.S. was partially supported by projects  MINECO
MTM2012-30951/FEDER, Gen. Cat. DGR2009SGR1040, and by ESF
EUROCORES program EuroGIGA-ComPoSe IP04-MICINN project
EUI-EURC-2011-4306.

\bibliographystyle {abbrv}
\bibliography {refs}
\normalsize

\clearpage
\appendix

\section{Detailed Proof of Lemma~\ref{lem:disks}}
\label {app:disks}

\begin{repeatlemma}{lem:disks}
Let $\mu$ be a Gaussian distribution with standard deviation $\sigma$. Let $k$ be a given integer. Then the minimum-error approximation of $\mu$ by a cylindrical step function $\mu_D$ consisting of $k$ disks is given by
\begin{equation}\label{eq:disksb}
\begin{aligned}
r_{i}&=\displaystyle \sqrt{2\sigma^{2}\log{\frac{k(k+1)}{(k+1-i)^{2}}}}\,,\\
w_i&=\displaystyle \frac{1}{\pi\sigma^2}\frac{(k+1-i)}{k (k+1)}\,,
\end{aligned}
\end{equation}
where $i\in(1,\dots,k)$.
\end{repeatlemma}

\begin{proof}
To find the optimal weights, first, write $W_i = \sum_{j=1}^i w_j$. We introduce an additional set of parameters $\rho_1, \ldots, \rho_k$, where $\rho_i$ is the radius such that $\mu(\rho_i) = W_i$, that is, it is those radii where the approximation and the true function intersect each other (see Figure~\ref{fig:graph}). We will optimize over the $2k$ variables $r_i$ and $\rho_i$, and derive the corresponding weights as a last step.
Now, let $D(\chi)$ be the complement of the open disk of radius $\chi$, centered at the origin.
Let $V(\chi)$ be the volume under the probability distribution $\mu$ in $D(\chi)$:
\[
V(\chi)=\iint\limits_{D(\chi)}{\mu(x,y)\dif x \dif y}=\int\limits_{0}^{2\pi}\int\limits_{\chi}^{\infty}{\mu(r,\theta)r}\dif r\dif\theta=\int\limits_{\chi}^{\infty}{\frac{r}{\sigma^2}e^{-\frac{r^2}{2\sigma^2}}}\dif r=e^{-\frac{\chi^2}{2\sigma^2}}\,.
\]

Then the symmetric difference between the function $\mu$ and $\mu_D$ is defined by $r_i$ and $\rho_i$ is given by the following equation:
\begin{equation}\label{eq:volume-dif}
\begin{split}
F&=  W_1(\pi r_{1}^{2}-\pi\rho_{1}^{2})-(V(\rho_{1})-V(r_1))+(V(r_{1})-V(\rho_{2}))-W_2(\pi\rho_{2}^{2}-\pi r_{1}^{2})\\
&\phantom{{}={}} +W_2(\pi r_{2}^{2}-\pi\rho_{2}^{2})-(V(\rho_{2})-V(r_{2}))+(V(r_{2})-V(\rho_{3}))-W_3(\pi\rho_{3}^{2}-\pi r_{3}^{2})\\
&\phantom{{}={}} +\dots\\
&\phantom{{}={}} +W_k(\pi r_{k}^{2}-\pi\rho_{k}^{2})-(V(\rho_{k})-V(r_{k}))+V(r_{k})\\
&=-V(\rho_{1})+2\sum_{i=1}^{k}{V(r_{i})}-2\sum_{i=2}^{k}{V(\rho_{i})}+\pi\sum_{i=1}^{k}{r_{i}^{2}\left(\mu(\rho_{i})+\mu(\rho_{i+1})\right)}-2\pi\sum_{i=2}^{k}{\rho_{i}^{2}}\mu(\rho_{i})\\
&=-1+2\sum_{i=1}^{k}{e^{-\frac{r_{i}^{2}}{2\sigma^{2}}}}-2\sum_{i=2}^{k}{e^{-\frac{\rho_{i}^{2}}{2\sigma^{2}}}}+\frac{1}{2\sigma^{2}}\sum_{i=1}^{k}{r_{i}^{2}\left(e^{-\frac{\rho_{i}^{2}}{2\sigma^{2}}}+e^{-\frac{\rho_{i+1}^{2}}{2\sigma^{2}}}\right)}-\frac{1}{\sigma^{2}}\sum_{i=2}^{k}{\rho_{i}^{2}}e^{-\frac{\rho_{i}^{2}}{2\sigma^{2}}}
\end{split}
\end{equation}
To minimize $F$, we compute the derivatives in $r_i$ and $\rho_i$, which leads to:
\[
\begin{aligned}
\frac{dF}{d\rho_{i}}
&=\frac{\rho_{i}}{\sigma^4}e^{-\frac{\rho_{i}^2}{2\sigma^2}}\left(\rho_{i}^2-\frac{1}{2}(r_{i-1}^2+r_{i}^2)\right)\,,\\
\frac{dF}{dr_{i}}
&=\frac{r_{i}}{\sigma^2}\left(e^{-\frac{\rho_{i}^2}{2\sigma^2}}+e^{-\frac{\rho_{i+1}^2}{2\sigma^2}}-2e^{-\frac{r_{i}^2}{2\sigma^2}}\right)\,.
\end{aligned}
\]

Setting the derivatives to $0$ results in the identities
\begin{align}
2\rho^2_{i}&=r^2_{i-1}+r^2_{i}\ \label{eq:eq1b},\\
2e^{-\frac{r^2_{i}}{2\sigma^2}}&=
\begin{cases}
\phantom{\Big|}e^{-\frac{\rho^2_{i}}{2\sigma^2}}+e^{-\frac{\rho^2_{i+1}}{2\sigma^2}} & \text{for }1\leq i<k\\
\phantom{\Big|}e^{-\frac{\rho^2_{i}}{2\sigma^2}} & \text{for }i=k\\
\end{cases}\,\label{eq:eq2b}\,.
\end{align}
To find the closed forms of expressions for $r_i$ and $\rho_i$, we do the following. First, if we substitute Equation~\ref{eq:eq1b} into Equation~\ref{eq:eq2b} we will get:
\[
e^{-\frac{r^2_{i}}{4\sigma^2}}=\frac{e^{-\frac{r^2_{i-1}}{4\sigma^2}}+e^{-\frac{r^2_{i+1}}{4\sigma^2}}}{2}\,.
\]
Define a function $g[i]=e^{-\frac{r^2_{i}}{4\sigma^2}}$, then the expression above can be rewritten as:
\[
g[i]=\frac{g[i-1]+g[i+1]}{2}\,.
\]
Notice, that this relation occurs only for linear functions, i.e.,
\[
g[i]=ai+b\,,
\]
for some coefficients $a$ and $b$. From Equation~\ref{eq:eq2b}, for $i=k$, we get
\[
2e^{-\frac{r^2_{k}}{2\sigma^2}}=e^{-\frac{r^2_{k-1}+r^2_{k}}{4\sigma^2}}\,,
\]
therefore
\[
g[k-1]=2g[k]\,.
\]
Using this equation we can express $a$ and $b$ as functions of $g[k]$, and get the following expression
\[
g[i]=(k+1-i)g[k]\,.
\]
Now, from Equation~\ref{eq:eq2b} for $i=1$ we get
\[
2e^{-\frac{r_{1}^2}{2\sigma^2}}=1+e^{-\frac{r_{1}^{2}+r_{2}^{2}}{4\sigma^2}}\,,
\]
therefore
\[
2g[1]^{2}=1+g[1]g[2]\,,
\]
and, finally,
\[
2k^{2}g[k]^{2}=1+k(k-1)g[k]^{2}\,.
\]
Therefore,
\[
g[k]=\frac{1}{\sqrt{k(k+1)}}\,,
\]
and
\[
e^{-\frac{r_{i}^{2}}{2\sigma^{2}}}\equiv g[i]^{2}=\frac{(k+1-i)^{2}}{k(k+1)}\,.
\]
Lastly, from this expression and Equation~\ref{eq:eq1b} we derive Equations~\ref{eq:disks} and the formula for $\rho_i$:
\begin{equation}
\rho_{i}=\sqrt{2\sigma^{2}\log{\frac{k(k+1)}{(k+1-i)(k+2-i)}}}\,.
\label{eq:rho}
\end{equation}

Substituting $w_i = W_i - W_{i+1} = \mu(\rho_i) - \mu(\rho_{i+1})$, we attain Equation~(\ref{eq:disksb}), proving the lemma.
\end{proof}

\section{Closed-Form Evaluation of Equation~(\ref{eq:subst-integral})}
\label{sec:appx-integral}
Here we'll show how to calculate the following integral for a cell $C$ of the partition $L^{*}$ of in the dual space:
\[
I=\iint\limits_{C} \left( \int\limits_{X_{1}(\alpha,\beta)}^{{X_{2}(\alpha,\beta)}} \int\limits_{X_{3}(\alpha,\beta)}^{{X_{4}(\alpha,\beta)}}(x_{2}-x_{1}) \dif{x_{2}} \dif{x_{1}}\right) \dif\alpha \dif\beta\,.
\]
Suppose lines corresponding to $C$ intersect four segments $s_{1},s_{2},s_{3}$, and $s_{4}$ that belong to the lines with the following equations:
\[
\begin{split}
a_{1}x+b_{1}y+c_{1}=0\,,&\quad a_{2}x+b_{2}y+c_{2}=0\,,\\
a_{3}x+b_{3}y+c_{3}=0\,,&\quad a_{4}x+b_{4}y+c_{4}=0\,.
\end{split}
\]
Then, the limits of integration can be expressed as:
\[
\begin{split}
X_{1}(\alpha,\beta)=\frac{b_{1}\beta-c_{1}}{b_{1}\alpha+a_{1}},\quad&X_{2}(\alpha,\beta)=\frac{b_{2}\beta-c_{2}}{b_{2}\alpha+a_{2}}\,,\\
X_{3}(\alpha,\beta)=\frac{b_{3}\beta-c_{3}}{b_{3}\alpha+a_{3}},\quad&X_{4}(\alpha,\beta)=\frac{b_{4}\beta-c_{4}}{b_{4}\alpha+a_{4}}\,.
\end{split}
\]
After solving the inner two integrals we get:
\[
\begin{split}
I=\iint\limits_{C}& \frac{(X_{2}\!-\!X_{1})(X_{4}\!-\!X_{3})(X_{3}\!+\!X_{4}\!-\!X_{1}\!-\!X_{2})}{2}\dif\alpha\dif\beta=\\
=\frac{1}{2}\iint\limits_{C}& \big(-X_{1}^{2} X_{3} + X_{2}^{2} X_{3} + X_{1} X_{3}^{2} - X_{2} X_{3}^{2}\\
&+X_{1}^{2} X_{4} - X_{2}^{2} X_{4} - X_{1} X_{4}^{2} + X_{2} X_{4}^{2}\big) \dif\alpha\dif\beta\,.
\end{split}
\]
For some $i$ and $j$:
\[
X_{i}X_{j}^{2}=\frac{(b_{i}\beta-c_{i})(b_{j}\beta-c_{j})^2}{(b_{i}\alpha+a_{i})(b_{j}\alpha+a_{j})^2}\,.
\]
Denote $I_{ij}$ to be:
\[
\begin{split}
I_{ij}=&\iint\limits_{C}X_{i}X_{j}^{2}\dif\alpha\dif\beta=\\
=&\iint\limits_{C}\frac{(b_{i}\beta-c_{i})(b_{j}\beta-c_{j})^2}{(b_{i}\alpha+a_{i})(b_{j}\alpha+a_{j})^2}\dif\alpha\dif\beta=\\
=&\sum_{C_{v}\subset C}\int\limits_{\alpha_{1}}^{\alpha_{2}} \left(\int\limits_{A_{1}\alpha+B_{1}}^{A_{2}\alpha+B_{2}} \frac{(b_{i}\beta-c_{i})(b_{j}\beta-c_{j})^2}{(b_{i}\alpha+a_{i})(b_{j}\alpha+a_{j})^2}\dif\beta\right)\dif\alpha\,,
\end{split}
\]
where cell $C$ is split into vertical splines $C_{v}$, with each $C_{v}$ bounded by left and right vertical segments with $\alpha$-coordinates equal to $\alpha_{1}$ and $\alpha_{2}$, and bottom and top segments defined by formulas $\beta=A_{1}\alpha+B_{1}$ and $\beta=A_{2}\alpha+B_{2}$. Then,
\[
I=\frac{1}{2}\left(I_{13}-I_{31}-I_{23}+I_{32}-I_{14}+I_{41}+I_{24}-I_{42}\right).
\]
Denote $F_{ij}(\alpha)$ to be an indefinite integral with additive constant equal to $0$:
\begin{equation}
F_{ij}(\alpha)=\int \left(\int\limits_{A_{1}\alpha+B_{1}}^{A_{2}\alpha+B_{2}} \frac{(b_{i}\beta-c_{i})(b_{j}\beta-c_{j})^2}{(b_{i}\alpha+a_{i})(b_{j}\alpha+a_{j})^2}\dif\beta\right)\dif\alpha\,.
\end{equation}
In the general case, when $b_{i}\not=0$, $b_{j}\not=0$, and $a_{i}/b_{i}\not=a_{j}/b_{j}$,
\[
\begin{split}
F_{ij}(\alpha)=&\frac{\log \left(a_i+\alpha b_i\right)}{12 b_i^2 \left(a_j b_i-a_i b_j\right)^2}\bigg[3 \left(A_2^4-A_1^4\right) a_i^4 b_j^2-12 a_i^3 \left(A_2^3 B_2-A_1^3 B_1\right) b_i b_j^2\\
&\qquad\qquad+18 a_i^2 \left(A_2^2 B_2^2-A_1^2 B_1^2\right) b_i^2 b_j^2-12 a_i \left(A_2 B_2^3-A_1 B_1^3\right) b_i^3 b_j^2+3 \left(B_2^4-B_1^4\right) b_i^4 b_j^2\\
&\qquad\qquad+ \left(b_j c_i+2 b_i c_j\right)\Big(4 \left(A_2^3-A_1^3\right) a_i^3 b_j-12 a_i^2 \left(A_2^2 B_2-A_1^2 B_1\right) b_i b_j \\
&\qquad\qquad\qquad\qquad+12 a_i \left(A_2 B_2^2-A_1 B_1^2\right) b_i^2 b_j -4 \left(B_2^3-B_1^3\right) b_i^3 b_j\Big)\\
&\qquad\qquad+\left(2 b_j c_i+b_i c_j\right)\Big(6 \left(A_2^2-A_1^2\right) a_i^2 b_i c_j -12 a_i \left(A_2 B_2-A_1 B_1\right) b_i^2 c_j +6 \left(B_2^2-B_1^2\right) b_i^3 c_j\Big)\\
&\qquad\qquad+12 \left(A_2-A_1\right) a_i b_i^2 c_i c_j^2-12 \left(B_2-B_1\right) b_i^3 c_i c_j^2
\bigg]\\
&+\frac{\log \left(a_j+\alpha  b_j\right)}{12 b_j^2 \left(a_j b_i-a_i b_j\right)^2}\bigg[3 \left(A_2^4-A_1^4\right) a_j^3 b_i \left(3 a_j b_i-4 a_i b_j\right)+12 a_j^2 \left(A_2^3 B_2-A_1^3 B_1\right) b_i b_j \left(3 a_i b_j-2 a_j b_i\right)\\
&\qquad\qquad+18 a_j \left(A_2^2 B_2^2-A_1^2 B_1^2\right) b_i b_j^2 \left(a_j b_i-2 a_i b_j\right)+12 a_i \left(A_2 B_2^3-A_1 B_1^3\right) b_i b_j^4-3 \left(B_2^4-B_1^4\right) b_i^2 b_j^4\\
&\qquad\qquad+\left(b_j c_i+2 b_i c_j\right)\Big(4 \left(A_2^3-A_1^3\right) a_j^2 \left(2 a_j b_i-3 a_i b_j\right)-12 a_j \left(A_2^2 B_2-A_1^2 B_1\right) b_j \left(a_j b_i-2 a_i b_j\right)\\
&\qquad\qquad\qquad\qquad-12 a_i \left(A_2 B_2^2-A_1 B_1^2\right) b_j^3+4 \left(B_2^3-B_1^3\right) b_i b_j^3\Big)\\
&\qquad\qquad+(2 b_j c_i+b_i c_j)\Big(6 \left(A_2^2-A_1^2\right) a_j c_j \left(a_j b_i-2 a_i b_j\right)+12 a_i \left(A_2 B_2-A_1 B_1\right) b_j^2 c_j-6 \left(B_2^2-B_1^2\right) b_i b_j^2 c_j\Big)\\
&\qquad\qquad-12 a_i b_j^2 c_i c_j^2 \left(A_2-A_1\right) -12 b_i b_j^2 c_i c_j^2 \left(B_2-B_1\right) 
\bigg]\\
&+\frac{1}{24 b_i b_j^2 \left(a_j b_i-a_i b_j\right) \left(a_j+\alpha  b_j\right)}\bigg[3\alpha^{3} b_i b_j^3 \left(a_j b_i-a_i b_j\right)\left(A_2^4-A_1^4\right)\\
&\qquad\qquad+\alpha^{2}\left(a_i b_j-a_j b_i\right)\Big(3b_j^2 \left(3 a_j b_i+2 a_i b_j\right)\left(A_2^4-A_1^4\right) - 24b_i b_j^3\left(A_2^3 B_2-A_1^3 B_1\right) \\
&\qquad\qquad\qquad\qquad +8 b_j^2 \left(b_j c_i+2 b_i c_j\right)\left(A_2^3-A_1^3\right)\Big)\\
&\qquad\qquad-\alpha\left(a_j b_i-a_i b_j\right)\Big(6 a_j b_j \left(2 a_j b_i+a_i b_j\right) \left(A_2^4-A_1^4\right) - 24 a_j b_i b_j^2\left(A_2^3 B_2-A_1^3 B_1\right)\\
&\qquad\qquad\qquad\qquad+8 a_j b_j\left(b_j c_i+2 b_i c_j\right)\left(A_2^3-A_1^3\right)\Big)\\
&\qquad\qquad+6 \left(A_2^4-A_1^4\right) a_j^4 b_i^2-24 a_j^3 \left(A_2^3 B_2-A_1^3 B_1\right) b_i^2 b_j\\
&\qquad\qquad+36 a_j^2 \left(A_2^2 B_2^2-A_1^2 B_1^2\right) b_i^2 b_j^2-24 a_j \left(A_2 B_2^3-A_1 B_1^3\right) b_i^2 b_j^3+6 \left(B_2^4-B_1^4\right) b_i^2 b_j^4\\
&\qquad\qquad+8 \left(A_2^3-A_1^3\right) a_j^3 b_i \left(b_j c_i+2 b_i c_j\right)-24 a_j^2 \left(A_2^2 B_2-A_1^2 B_1\right) b_i b_j \left(b_j c_i+2 b_i c_j\right)\\
&\qquad\qquad+24 a_j \left(A_2 B_2^2-A_1 B_1^2\right) b_i b_j^2 \left(b_j c_i+2 b_i c_j\right)-8 \left(B_2^3-B_1^3\right) b_i b_j^3 \left(b_j c_i+2 b_i c_j\right)\\
&\qquad\qquad+12 \left(A_2^2-A_1^2\right) a_j^2 b_i c_j \left(2 b_j c_i+b_i c_j\right)-24 a_j \left(A_2 B_2-A_1 B_1\right) b_i b_j c_j \left(2 b_j c_i+b_i c_j\right)\\
&\qquad\qquad+12 \left(B_2^2-B_1^2\right) b_i b_j^2 c_j \left(2 b_j c_i+b_i c_j\right)+24 \left(A_2-A_1\right) a_j b_i b_j c_i c_j^2-24 \left(B_2-B_1\right) b_i b_j^2 c_i c_j^2\bigg]\,.
\end{split}
\]
If segment $i$ is pointing towards segment $j$ (line drawn through $i$ intersects $j$), and one of the corners of the integration spline corresponds to the line going through $i$, then the following equalities hold
\[
\begin{split}
a_i+\alpha' b_i&=0\,,\\
A_1a_i-B_1b_i+c_i&=0\,,\\
A_2 a_i-B_2 b_i+c_i&=0\,,\\
\end{split}
\]
where $\alpha'$ corresponds to the corner of the spline. In that case,
\[
\begin{split}
F_{ij}(\alpha)=&\frac{\log \left(a_j+\alpha  b_j\right)}{12 b_i^2 b_j^2}\bigg[9 \left(A_2^4-A_1^4\right) \left(a_j b_i-a_i b_j\right)^2+16 \left(A_2^3-A_1^3\right) \left(a_j b_i-a_i b_j\right) \left(b_i c_j-b_j c_i\right)\\
&\qquad\qquad+6 \left(A_2^2-A_1^2\right) \left(b_j c_i-b_i c_j\right)^2\bigg]\\
&+\frac{1}{24 b_i^3 b_j^2 \left(a_j+\alpha  b_j\right)}\bigg[3\alpha^{4}b_i^3 b_j^3\left(A_2^4-A_1^4\right)\\
&\qquad\qquad-\alpha^{2}b_i^2 b_j^2\Big(9 \left(a_j b_i-2 a_i b_j\right)\left(A_2^4-A_1^4\right)+16 \left(b_i c_j-b_j c_i\right)\left(A_2^3-A_1^3\right)\Big)\\
&\qquad\qquad+2a_j b_i^2 b_j\alpha\Big(3\left(3 a_i b_j-2 a_j b_i\right)\left(A_2^4-A_1^4\right)-8 \left(b_i c_j-b_j c_i\right)\left(A_2^3-A_1^3\right)\Big)\\
&\qquad\qquad+\left(6 a_j^2 b_i^2-9 a_i^2 b_j^2\right)\left(A_2^4-A_1^4\right)+16 a_j b_i \left(b_i c_j-b_j c_i\right)\left(A_2^3-A_1^3\right)+12\left(b_j c_i-b_i c_j\right)^2\left(A_2^2-A_1^2\right)\bigg]\,.
\end{split}
\]
If segment $j$ is pointing towards segment $i$ (line drawn through $j$ intersects $i$), and one of the corners of the integration spline corresponds to the line going through $j$, then the following equalities hold
\[
\begin{split}
a_j+\alpha' b_j&=0\,,\\
A_1a_j-B_1b_j+c_j&=0\,,\\
A_2 a_j-B_2 b_j+c_j&=0\,,\\
\end{split}
\]
where $\alpha'$ corresponds to the corner of the spline. In that case,
\[
\begin{split}
F_{ij}(\alpha)=&\frac{\log \left(a_i+\alpha  b_i\right)}{12 b_i^2 b_j^2}\bigg[4 \left(A_2^3-A_1^3\right) \left(a_j b_i-a_i b_j\right) \left(b_i c_j-b_j c_i\right)+3 \left(A_2^4-A_1^4\right) \left(a_j b_i-a_i b_j\right)^2\bigg]\\
&+\frac{1}{24 b_i^2 b_j^2}\bigg[3 \alpha^{2}\left(A_2^4-A_1^4\right) b_i b_j+\alpha\Big(\left(A_2^4-A_1^4\right) \left(12 a_j b_i-6 a_i b_j\right)-8 \left(A_2^3-A_1^3\right) \left(b_j c_i-b_i c_j\right)\Big)\bigg]\,.
\end{split}
\]
In case when $b_{i}\not=0$, $b_{j}\not=0$, but lines are parallel ($a_{i}/b_{i}=a_{j}/b_{j}$),
\[
\begin{split}
F_{ij}(\alpha)=&\frac{\log \left(a_i+\alpha b_i\right)}{2 b_i^2 b_j^2}\bigg[
3 \left(A_2^4-A_1^4\right) a_i^2 b_j^2-6 a_i \left(A_2^3 B_2-A_1^3 B_1\right) b_i b_j^2+3 \left(A_2^2 B_2^2-A_1^2 B_1^2\right) b_i^2 b_j^2\\
&\qquad\qquad+2 \left(A_2^3-A_1^3\right) a_i b_j \left(b_j c_i+2 b_i c_j\right)-2 \left(A_2^2 B_2-A_1^2 B_1\right) b_i b_j \left(b_j c_i+2 b_i c_j\right)\\
&\qquad\qquad+\left(A_2^2-A_1^2\right) b_i c_j \left(2 b_j c_i+b_i c_j\right)
\bigg]\\
&+\frac{1}{24 b_i^2 b_j^2 \left(a_i+\alpha b_i\right)^2}\bigg[3 \alpha^{4} \left(A_1^4-A_2^4\right) b_i^4 b_j^2\\
&\qquad\qquad+\alpha^{3}\Big(12 a_i b_i^3 b_j^2(A_2^4-A_1^4)-24 b_i^4 b_j^2(A_2^3 B_2-A_1^3 B_1)+8 b_i^3 b_j \left(b_j c_i+2 b_i c_j\right)(A_2^3-A_1^3)\Big)\\
&\qquad\qquad+\alpha^{2}\Big(33 a_i^2 b_i^2 b_j^2(A_2^4-A_1^4)-48 a_i b_i^3 b_j^2(A_2^3 B_2-A_1^3 B_1)+16 a_i b_i^2 b_j \left(b_j c_i+2 b_i c_j\right)(A_2^3-A_1^3)\Big)\\
&\qquad\qquad-\alpha\Big(6 a_i^3 b_i b_j^2(A_2^4-A_1^4)-48 a_i^2 b_i^2 b_j^2(A_2^3 B_2-A_1^3 B_1)+72 a_i b_i^3 b_j^2(A_2^2 B_2^2-A_1^2 B_1^2)\\
&\qquad\qquad\qquad\qquad-24 b_i^4 b_j^2(A_2 B_2^3-A_1 B_1^3)+16 a_i^2 b_i b_j \left(b_j c_i+2 b_i c_j\right)\left(A_2^3-A_1^3\right)\\
&\qquad\qquad\qquad\qquad-48 a_i b_i^2 b_j \left(b_j c_i+2 b_i c_j\right)\left(A_2^2 B_2-A_1^2 B_1\right)+24 b_i^3 b_j \left(b_j c_i+2 b_i c_j\right)\left(A_2 B_2^2-A_1 B_1^2\right)\\
&\qquad\qquad\qquad\qquad+24 a_i b_i^2 c_j \left(2 b_j c_i+b_i c_j\right)\left(A_2^2-A_1^2\right)-24 b_i^3 c_j \left(2 b_j c_i+b_i c_j\right)\left(A_2 B_2-A_1 B_1\right)\\
&\qquad\qquad\qquad\qquad+24 b_i^3 c_i c_j^2 \left(A_2-A_1\right)\Big)\\
&\qquad\qquad-21 a_i^4 b_j^2\left(A_2^4-A_1^4\right)+60 a_i^3 b_i b_j^2\left(A_2^3 B_2-A_1^3 B_1\right)-54 a_i^2 b_i^2 b_j^2\left(A_2^2 B_2^2-A_1^2 B_1^2\right)\\
&\qquad\qquad+12 a_i b_i^3 b_j^2\left(A_2 B_2^3-A_1 B_1^3\right)+3 b_i^4 b_j^2 \left(B_2^4-B_1^4\right)\\
&\qquad\qquad-\left(b_j c_i+2 b_i c_j\right)\Big(20 a_i^3 b_j \left(A_2^3-A_1^3\right) -36 a_i^2 b_i b_j\left(A_2^2 B_2-A_1^2 B_1\right)\\
&\qquad\qquad\qquad\qquad+12 a_i b_i^2 b_j \left(A_2 B_2^2-A_1 B_1^2\right)+4 b_i^3 b_j\left(B_2^3-B_1^3\right)\Big)\\
&\qquad\qquad-\left(2 b_j c_i+b_i c_j\right)\Big(18 a_i^2 b_i c_j\left(A_2^2-A_1^2\right)-12 a_i b_i^2 c_j\left(A_2 B_2-A_1 B_1\right)-6 b_i^3 c_j\left(B_2^2-B_1^2\right) \Big)\\
&\qquad\qquad-12 a_i b_i^2 c_i c_j^2\left(A_2-A_1\right)-12 b_i^3 c_i c_j^2 \left(B_2-B_1\right)\bigg]\,.
\end{split}
\]
If $b_{i}=0$ (segment $i$ is vertical), and $b_{j}\not=0$, then
\[
\begin{split}
F_{ij}(\alpha)=&\frac{\log \left(a_j+\alpha  b_j\right)}{a_i b_j^2}\bigg[-a_j^2 c_i\left(A_2^3-A_1^3\right) +2 a_j b_j c_i\left(A_2^2 B_2-A_1^2 B_1\right)- b_j^2 c_i\left(A_2 B_2^2-A_1 B_1^2\right)\\
&\qquad\qquad-2 a_j c_i c_j\left(A_2^2-A_1^2\right)+2 b_j c_i c_j\left(A_2 B_2-A_1 B_1\right)-c_i c_j^2\left(A_2-A_1\right) \bigg]\\
&+\frac{1}{6 a_i b_j^2 \left(a_j+\alpha  b_j\right)}\bigg[-\alpha^{3}b_j^3 c_i\left(A_2^3-A_1^3\right)\\
&\qquad\qquad+3\alpha b_j^2 c_i\Big(a_j\left(A_2^3-A_1^3\right)+2 b_j \left(A_1^2 B_1-A_2^2 B_2\right) +2 c_j\left(A_2^2-A_1^2\right)\Big)\\
&\qquad\qquad+2\alpha a_j b_j c_i\Big(2 a_j\left(A_2^3-A_1^3\right)+3 b_j\left(A_1^2 B_1-A_2^2 B_2\right) +3c_j\left(A_2^2-A_1^2\right)\Big)\\
&\qquad\qquad-2 a_j^3 c_i\left(A_2^3-A_1^3\right)+6 a_j^2 b_j c_i\left(A_2^2 B_2-A_1^2 B_1\right)-6 a_j b_j^2 c_i\left(A_2 B_2^2-A_1 B_1^2\right)+2 b_j^3 c_i\left(B_2^3-B_1^3\right)\\
&\qquad\qquad-6 a_j^2 c_i c_j\left(A_2^2-A_1^2\right)+12 a_j b_j c_i c_j\left(A_2 B_2-A_1 B_1\right)-6 b_j^2 c_i c_j\left(B_2^2-B_1^2\right)\\
&\qquad\qquad-6 a_j c_i c_j^2\left(A_2-A_1\right)+6 b_j c_i c_j^2\left(B_2-B_1\right)\bigg]\,.
\end{split}
\]
If segment $j$ is pointing towards segment $i$:
\[
\begin{split}
a_j+\alpha' b_j&=0\,,\\
A_1a_j-B_1b_j+c_j&=0\,,\\
A_2 a_j-B_2 b_j+c_j&=0\,,\\
\end{split}
\]
where $\alpha'$ corresponds to the corner of the spline. In that case,
\[
F_{ij}(\alpha)=\frac{\alpha c_i \left(2 a_j+\alpha  b_j\right)\left(A_1^3-A_2^3\right)}{6 a_i b_j}\,.
\]
If $b_{i}\not=0$, and $b_{j}=0$ (segment $j$ is vertical), then
\[
\begin{split}
F_{ij}(\alpha)=&\frac{\log \left(a_i+\alpha  b_i\right)}{2 a_j^2 b_i^2}\bigg[a_i^2 c_j^2\left(A_2^2-A_1^2\right)-2 a_i b_i c_j^2\left(A_2 B_2-A_1 B_1\right)+ b_i^2 c_j^2\left(B_2^2-B_1^2\right)\\
&\qquad\qquad+2 a_i c_i c_j^2\left(A_2-A_1\right)-2b_i c_i c_j^2\left(B_2-B_1\right)  \bigg]\\
&+\frac{1}{6 a_i b_j^2 \left(a_j+\alpha  b_j\right)}\bigg[\alpha^{2}\left(A_2^2-A_1^2\right) b_i c_j^2\\
&\qquad\qquad-2\alpha c_j^2\Big(a_i\left(A_2^2-A_1^2\right) -2 b_i\left(A_2 B_2-A_1 B_1\right) +2 c_i\left(A_2-A_1\right)\Big)
\bigg]\,.
\end{split}
\]
If segment $i$ is pointing towards segment $j$:
\[
\begin{split}
a_i+\alpha' b_i&=0\,,\\
A_1a_i-B_1b_i+c_i&=0\,,\\
A_2 a_i-B_2 b_i+c_i&=0\,,\\
\end{split}
\]
where $\alpha'$ corresponds to the corner of the spline. In that case,
\[
F_{ij}(\alpha)=\frac{\alpha c_j^2 \left(2 a_i+\alpha b_i\right)\left(A_2^2-A_1^2\right)}{4 a_j^2 b_i}\,.
\]
If $b_{i}=0$ and $b_{j}=0$ (both segments are vertical), then
\[
F_{ij}(\alpha)=-\frac{c_i c_j^2 \left(\alpha^{2} (A_2-A_1)+2 \alpha(B_2- B_1)\right)}{2 a_i a_j^2}\,.
\]
\end {document}